\documentclass[aps,prl,twocolumn,showpacs,10pt,superscriptaddress,preprintnumbers,nofootinbib]{revtex4-2}
\usepackage{amsmath,bm}
\usepackage{physics}
\usepackage{graphicx}
\usepackage{amssymb}
\usepackage[colorlinks,citecolor=red]{hyperref}
\usepackage[normalem]{ulem}

\newcommand{\beq}{\begin{equation}}
\newcommand{\eeq}{\end{equation}}
\newcommand{\nn}{\nonumber}
\newcommand{\ivfb}{{\rm fb}^{-1}}

\begin{document}

\preprint{
	{\vbox {			
		\hbox{\bf MSUHEP-21-026}
}}}
\vspace*{0.2cm}

\title{Azimuthal Angular Correlation as a Boosted Top Jet Substructure}

\author{Zhite Yu}
\email{yuzhite@msu.edu}
\affiliation{Department of Physics and Astronomy,
Michigan State University, East Lansing, Michigan 48824, USA}

\author{C.-P. Yuan}
\email{yuanch@msu.edu}
\affiliation{Department of Physics and Astronomy,
Michigan State University, East Lansing, Michigan 48824, USA}

\date{\today}

\begin{abstract}
We propose a novel jet substructure observable of boosted tops that is related to the linear polarization of the $W$ boson in boosted top quark decay, which results in a $\cos2\phi$ angular correlation between the $t\to bW$ and $W\to f\bar{f'}$ decay planes. 
We discuss in detail the origin of such linear polarization by applying Wigner's little group transformation. We show that the unique $\cos2\phi$ angular correlation only exists in the boosted regime but not in the top quark rest frame. We construct an experimental observable for such correlation based on the transverse energy deposition asymmetry in the top jet that does not require the reconstruction of $W$ decay products. The degree of this asymmetry can be used to measure the longitudinal polarization of the top quark, which is an important probe of new physics that couples to the top sector, and can discriminate a boosted top quark jet from its background events, such as QCD jets. A numerical simulation is also performed and found to agree well with the analytic prediction of the Standard Model.
\end{abstract}

\maketitle

\emph{Introduction.}---Boosted top quarks, with their energies much greater than their mass, 
provide a unique opportunity for testing the Standard Model (SM) and searching for new physics (NP)~\cite{Schatzel:2013wsr}. In this kinematic region, the top quark decay products are collimated, resembling a light QCD jet in appearance. Such a cone signature enhances the selection efficiency of boosted top quark events with respect to the background, which compensates for the small production rate~\citep{Abdesselam:2010pt}. In addition, the semileptonic decay mode no longer possesses special advantage over the hadronic mode, and one ought to take the latter into account to enhance the statistics. Then, the boosted top can be readily identified as a single ``fat" jet by some jet algorithm and becomes difficult to distinguish from a QCD jet. Hence, for the experimental study of boosted tops, one needs first to be able to distinguish a boosted top quark jet from a QCD jet.

There have been many tagging algorithms proposed and applied to discriminate boosted top quark events from QCD jets~\cite{CMS-PAS-JME-13-007, Plehn:2010st, ATLAS:2018wis}.
Also, machine learning methods have been applied and found to improve the tagging efficiency substantially, especially when multiple taggers are included~\cite{10.21468/SciPostPhys.7.1.014, Bhattacharya:2020aid, arXiv:2205.02817}.
Those techniques mainly make use of the top and $W$ mass conditions and the three-subjet structure.
In this Letter, we propose a new substructure observable of the boosted top quark jet that exploits the azimuthal angular correlation among the decay products without the need to identify the two-pronged decay signature of the $W$ boson. When used together with other top taggers, this observable shall further improve the tagging efficiency.

\begin{figure}[htbp]
	\centering
	\includegraphics[scale=0.5]{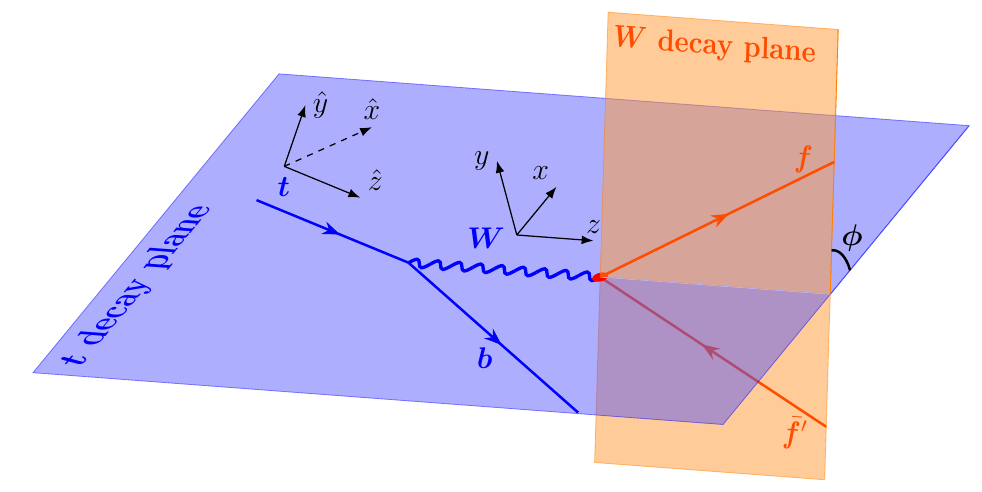}
	\caption{The two successive decay planes in $t\to bW (\to f\bar{f}')$ decay process. The coordinate systems of top frame and $W$ frame are shown separately. The $x$ axis of $W$ frame lies on the $t$ decay plane, while the $\hat{x}$ axis of the top frame may not.}
	\label{fig:frame}
\end{figure}

The azimuthal correlation of interest to us is the angle between the decay planes of $t\to b W$ and $W\to f \bar{f}'$, shown as the $\phi$ angle in Fig.~\ref{fig:frame}. 
We point out an interesting $\cos2\phi$ distribution that arises from the linear polarization of $W$, which is a superposition of its $+1$ and $-1$ helicity eigenstates. Such polarization does not exist in the top rest frame but emerges as a result of $W$ helicity mixing when going to the boosted top frame, which makes it a unique observable for the boosted top jet. 
We will show that this phenomenon is generic to many boosted $1\to 3$ decay systems, especially for QCD jets~\cite{Chen:2020adz}. 
Nevertheless, the correlation in QCD jets is much less significant than that in the boosted top jet, so the azimuthal correlation can be used as a top tagger against QCD jets.

The measurement of top quark polarization is important for testing the SM and exploring NP models~\cite{Kane:1991bg,Berger:2011hn}, which is commonly done in the top rest frame for the semileptonic decay mode~\cite{ATLAS-CONF-2021-027, ATLAS:2013gil, Jezabek:1994qs, Brandenburg:2002xr, CMS:2019nrx, Mahlon:2010gw, Schwienhorst:2010je, Aguilar-Saavedra:2017wpl}. In the boosted regime, however, it is no longer a good choice to go to the top rest frame because the finite granular size of the detector leads to large uncertainties of the  angular separations (especially in polar angles) among the subjets inside the top jet, inhibiting the full reconstruction of its rest frame. As a result, we study the boosted top polarization within the boosted regime.
For this purpose, some methods have been designed~\cite{Shelton:2008nq, Krohn:2009wm, Kitadono:2015nxf, Godbole:2019erb} by making use of the energy or polar 
angular distribution of the decay products. Below, we will show 
how the $\cos2\phi$ azimuthal correlation can serve as an additional top polarimeter in the boosted regime, and we will propose an experimental observable for extracting this correlation.

\emph{$W$ density matrix.}---In a top quark production process, we choose the $\hat{z}$ axis as the top moving direction, while the $\hat{x}$ axis lies on its production plane and $\hat{y} = \hat{z} \times \hat{x}$.
For instance, for a top production event at the LHC, the production plane is formed by the initial-state proton beams and the top momentum, and $\hat{y}$ is perpendicular to this plane.
In this frame, called the ``boosted top frame", the top is boosted with energy $E_t$. Its decay into $bf\bar{f'}$ can be described as two successive steps: first, $t$ decays to $b$ and $W$, whose polarization is described by the unnormalized density matrix
\beq
W_{\lambda_w\lambda_w'}
= \rho^t_{\lambda_t\lambda'_t}
\mathcal{M}_{\lambda_t\lambda_w\lambda_b}
\mathcal{M}^*_{\lambda'_t\lambda'_w\lambda_b},
\label{eq:unnor W denmtx}
\eeq
where a sum over repeated indices is implied; and then $W$ decays into a 
fermion pair $f\bar{f'}$. $\mathcal{M}_{\lambda_t\lambda_w\lambda_b}$ is the helicity amplitude of $t\to bW$, and $\rho^t = \left(1+\bm{s}_t\cdot \bm{\sigma}\right)/2$ is the top quark's spin density matrix, with $\bm{s}_t = \left( b_1, b_2, h_t \right)$ being its polarization vector and $\bm{\sigma} = \left( \sigma_1, \sigma_2, \sigma_3 \right)$ the Pauli matrices.

Before proceeding with our discussion, we first define some frames. Apart from the boosted top frame already defined, 
we define the ``top rest frame" as having the {\it same} $\hat{x}$-$\hat{y}$-$\hat{z}$ coordinate system as the boosted top frame but with the top at rest, and we also define the ``$W$ frame" by having the $z$ axis along $\bm{p}_W$, and $y$ axis along $\bm{p}_b \times \bm{p}_W$, where $\bm{p}_b$ and $\bm{p}_W$ are the three-momenta of $b$ and $W$, respectively, in the boosted top frame. The ``$W$ rest frame" is obtained by boosting the $W$ frame back along $z$.
See Fig.~\ref{fig:frame} for a graphic illustration.

As a massive spin-1 particle, the $W$'s density matrix [Eq.~\eqref{eq:unnor W denmtx}] is a $3\times 3$ Hermitian matrix and so can be described by eight real parameters in addition to its trace. 
In the helicity basis ($W_+,W_L,W_-$), for $\lambda_w=1,0,-1$, it can be written as
\begin{widetext}
\begin{align}
\label{eq:W denmtx}
\left( W_{\lambda_w \lambda_w'} \right) =
\begin{pmatrix}
\frac{{\rm tr}W}{3} -\frac{\delta_L}{3} + \frac{J_z}{2} &
\frac{J_x + 2 Q_{xz} - i(J_y + 2 Q_{yz})}{2\sqrt{2}} &
\xi - i Q_{xy}		\\
\frac{J_x + 2 Q_{xz} + i(J_y + 2 Q_{yz})}{2\sqrt{2}} &
\frac{{\rm tr}W}{3}+\frac{ 2\delta_L}{3}	&
\frac{J_x - 2 Q_{xz} - i(J_y - 2 Q_{yz})}{2\sqrt{2}}  \\
\xi + i Q_{xy} &
\frac{J_x - 2 Q_{xz} + i(J_y - 2 Q_{yz})}{2\sqrt{2}} &
\frac{{\rm tr}W}{3}- \frac{\delta_L}{3} - \frac{J_z}{2}
\end{pmatrix},
\end{align}
\end{widetext}
where ${\rm tr}W$ is the production rate of $W$ boson,
$J_i$ is its spin angular momentum along the $i$th direction ($i=x,y,z$), and the others its quadrupole moments. They will be referred to as (unnormalized) $W$ polarization parameters. The diagonal elements describe the rates of each $W$ helicity state, and the off-diagonal ones arise from the interference between different helicity states. 

How does the azimuthal distribution depend on the $W$ polarization parameters?
If the $W$ boson is at helicity eigenstate $| \lambda_w \rangle$, the $\phi$ dependence of its decay products is fully captured by a phase factor $e^{i\lambda_w \phi}$, which ends up being a constant in the amplitude square. To get a nontrivial azimuthal dependence requires the {\it interference} between different helicity states. Among the polarization parameters in Eq.~\eqref{eq:W denmtx}, $\left( J_x,  Q_{xz}\right)$ and $\left( J_y,  Q_{yz}\right)$ are associated with $\cos\phi$ and $\sin\phi$ distributions, respectively, as they are the interference between $W_{\pm}$ and $W_L$ states, and $\xi$ and $Q_{xy}$ are associated with $\cos2\phi$ and $\sin2\phi$ modulations, respectively, for being the interference between $W_{+}$ and $W_-$ states.

The angular correlation between the two decay planes in the boosted top system manifests itself as a $\cos2\phi$ modulation in the SM.
We interpret this modulation in the $W$'s linear polarization basis, which consists of the states $\{ |x\rangle, |y\rangle, |z\rangle \}$, related to the helicity eigenstates by $|\pm\rangle = \left( \mp |x\rangle - i |y \rangle \right)/\sqrt{2}$ and $|0\rangle = |z \rangle$. In this basis, $\xi = \left( W_{yy} - W_{xx} \right) / 2$,
which means that $\xi$ denotes the difference between the fraction of linearly polarized $W$ events along $y$ and along $x$. The linear polarization sets a special azimuthal direction, which breaks the azimuthal rotational invariance in $W$'s decay so that the $f\bar{f}'$ plane tends to be perpendicular to the linear polarization of the $W$ boson. 
For example, if the $W$ were purely linearly polarized along $y$, the $f\bar{f}'$ plane would tend to be aligned with the $x$-$z$ plane; cf. Fig.~\ref{fig:frame}.

\emph{Origin of $\xi$.}---The specific values of $W$ density matrix $W_{\lambda_w \lambda_w'}$, and hence the polarization parameters, depend on the reference frame. In the top rest frame, the $t\to bW$ helicity amplitudes are constrained by angular momentum conservation.
Because $t$ has spin $1/2$, a certain $b$ helicity state cannot be produced together with both the $W_+$ and $W_-$ states. This conclusion holds even when the $b$ quark mass is not neglected.
For example, if $\lambda_b = -1/2$,
$W$ can only have $\lambda_w = -1$ or $0$ because $\lambda_w = +1$ would lead to a total spin momentum $3/2$ along the $W$ momentum direction, which cannot be produced from a spin-$1/2$ $t$.
So there cannot be any interference between the $W_+$ and $W_-$ states.
Consequently, in the top rest frame, $\xi$ must vanish, and hence there is no $\cos2\phi$ angular correlation.

Now, we go from the top rest frame to the boosted top frame by boosting the $tbW$ system along the $\hat{z}$ direction by the Lorentz boost transformation $\Lambda_t = \Lambda_z(\beta_t)$,
where $\beta_t = p_t / E_t$ is determined by the top momentum in the boosted top frame.
Under this boost, the $W$ helicity state transforms according to its little group~\cite{Weinberg:1995mt}, which is a rotation around the $\hat{y}$ axis by angle $\chi \in [0, \pi]$, with
\begin{align}
\cos\chi &=
\frac{v_w + \beta_t \cos\theta_w}{
\sqrt{ (1+\beta_t v_w \cos\theta_w)^2 - (1-\beta_t^2)(1-v_w^2) }
},
\label{eq:boost angle}
\end{align}
where $v_w$
and $\theta_w$ are, respectively,  the speed and polar angle of $W$ in the top rest frame.
The $W$ density matrix [Eq.~\eqref{eq:unnor W denmtx}] transforms as a rank-2 tensor by the Wigner-$d^1$ function, $\left( W_{\lambda\lambda'} \right) \to d^1(\chi) \cdot \left( W_{\lambda\lambda'} \right) \cdot [d^1(\chi)]^T$.
This leads to a mixing among $\xi$, $\delta_L$, and $Q_{xz}$, particularly with
\begin{align}
\xi' & = \frac{3\xi - \delta_L}{4} + 
\frac{1}{2}\left( 
Q_{xz} \sin2\chi + \frac{\xi +\delta_L}{2} \cos2\chi 
\right),
\label{eq:mixing}
\end{align}
where the primed (unprimed) polarization parameters refer to the ones in the boosted top frame (top rest frame).
Though we have shown that $\xi = 0$ in the top rest frame, a nonzero value ({\it i.e.}, $\xi'\neq 0$) can be generated in the boosted top frame due to the mixing.
The mixing originates from the massiveness of the $W$ boson and is the source of such {\it new} kind of polarization in the boosted top system that is absent in the top rest frame.  

It should be noted that the presence of $\cos2\phi$ modulation in the boosted top frame 
arises as a mixing with other nonzero parameters ($Q_{xz}$ and $\delta_L$) present in the top rest frame. As a whole, 
the physical information is conserved in both reference frames; it merely appears in a different form as a new $\cos2\phi$ distribution in the boosted top frame. Nevertheless, the $\cos2\phi$ distribution does have some advantages over the angular functions associated with $Q_{xz}$ and $\delta_L$, which are $\sin2\theta^\star_f \cos\phi_f$ and $(1-3\cos^2\theta^\star_f)/3$, respectively.
To measure the latter two angular distributions, it is necessary to both distinguish $f$ from $\bar{f}'$ and measure the polar angle ($\theta^\star_f$) of $f$ (or $\bar{f}'$) in the $W$ rest frame. Because of the finite angular resolution of the detector, it may become difficult to measure the polar angle precisely 
in the boosted case in order to reconstruct the $W$ rest frame. In contrast, the azimuthal angle is relatively easier to measure, since it only concerns the relative orientation of the energy deposits, and, due to the invariance of $\cos2\phi$ under $\phi \to \phi + \pi$, it only cares about the {\it plane} of $Wf\bar{f}'$ and does not require one to distinguish $f$ from $\bar{f}'$; the latter feature is important for detecting the boosted top quark in its hadronic decay mode.

\emph{Azimuthal angular correlation.}---Assuming the SM $W$-$t$-$b$ coupling, the azimuthal angular correlation between the fermion pair plane and the $bW$ plane in the boosted top quark jet takes the form
\begin{align}
P_t(\phi) \equiv 
\frac{\pi}{\Gamma_t}\frac{\dd{\Gamma_t}}{\dd\phi}
= 1 + \langle \xi' \rangle \cos2\phi ,
\quad
\phi \in [0, \pi),
\label{eq:phi distribution}
\end{align}
where $\langle \xi' \rangle \equiv \left( m_t^2 / 2 m_w^2 + 1\right)^{-1} \left( \int \dd{\Omega_w^{\star}} \xi' /  4 \pi \right)$
is the average of $\xi'$ over the $W$ angles. Since the angle between the two decay planes does not require one to distinguish $f$ from $\bar{f}'$, the above correlation can be measured in the hadronic decay mode of the top quark with $\phi \in [0, \pi)$.
To measure the above correlation in the  
semileptonic decay mode of the top quark, one needs to first reconstruct the missing neutrino three-momentum by imposing kinematic constraints of the event~\cite{ATLAS:2013nki, CMS:2012jea}.
In that case, one can use the full information of $\phi_f \in [0, 2\pi)$ to include an additional $\cos\phi_f$ angular dependence associated with the polarization parameter $J_x'$. 
Here, by focusing on the angle $\phi$ between the two decay planes, instead of the azimuthal angle $\phi_f$ of one particular particle from the $W$ decay, we only have to consider the $\cos2\phi$ angular correlation.

The coefficient $\langle \xi' \rangle$ depends on the top quark's energy $E_t$ and longitudinal polarization $h_t$ and takes the analytic form
\beq
\langle \xi' \rangle = \kappa(\beta_t, r) \cdot \left( h_t - \beta_t \right),
\label{eq:xi analytic}
\eeq
where $r = m_w/m_t$ and the spin analyzing power is 
\begin{widetext}
\begin{align}
\kappa(\beta_t, r)
= &
\frac{r}{2 \beta_t^{2} \sqrt{1-\beta_t^{2}}\left(1-r^{2}\right)^{2}\left(1+2 r^{2}\right)} 
\cdot
\left\{
4 r \sqrt{1-\beta_t^{2}} 
\left[ \left(1+r^{2}\right) \log \left|\frac{\beta_t\left(1+r^{2}\right)+\left(1-r^{2}\right)}{\beta_t\left(1+r^{2}\right)-\left(1-r^{2}\right)}\right| - \beta_t\left(1-r^{2}\right)\right]
\right. \nn\\
&\hspace{6em}
\left. - 
\left[ 4 r^{2}+\left(1-\beta_t^{2}\right)\left(1+r^{2}\right)^{2} \right] 
\tanh^{-1} \left[\frac{4 \beta_t \sqrt{1-\beta_t^{2}} \, r\left(1-r^{2}\right)}{\left(1-\beta_t^{2}\right)\left(1-r^{2}\right)^{2}+4 \beta_t^{2} r^{2}}\right]
\right\}.
\end{align}
\end{widetext}
The dependence on $E_t$ converges very quickly to the infinitely boosted limit, such that a top quark with $E_t \gtrsim 500~{\rm GeV}$ can already be considered as highly boosted.
Therefore, for phenomenological study of boosted tops, we can well approximate $\langle \xi' \rangle$ by its limit with $E_t = \infty$, which takes the numerical form $\langle \xi' \rangle \simeq 0.145 \left( h_t  - 1 \right)$,
with a spin-analyzing power 0.145. 

In the case for antitop quark, we have the same $\cos2\phi$ correlation as in Eq.~\eqref{eq:phi distribution}, but the coefficient $\langle \bar{\xi}' \rangle$ differs from Eq.~\eqref{eq:xi analytic} by $h_t \to -h_t$ due to {\it CP} invariance.

\emph{Comparison to QCD jet.}---The derivation of Eqs.~\eqref{eq:mixing} and \eqref{eq:phi distribution} makes it clear that the $\cos2\phi$ azimuthal correlation is not only relevant to boosted top quarks, but also to any boosted $1\to 3$ decay systems as long as they are mediated by virtual vector bosons, such as boosted QCD jets with a virtual gluon, boosted $b\to s l^+ l^-$ decay through a virtual photon or $Z$ boson, or $b\to c \bar{\nu}_l l^- $ decay via a virtual $W$. In more general cases with {\it CP} violation, there will also be an additional $\sin2\phi$ correlation.

A particular example is the three-pronged QCD jets, for which the azimuthal angular correlation $P_j(\phi) = 1 + \langle \xi_j \rangle \cos2\phi$ has been pointed out for the three-point energy correlator~\cite{Chen:2020adz}. This is relevant to boosted top quarks because QCD jets can be a source of background of the latter and needs to be distinguished when studying the hadronically decayed boosted top quarks.
However, there are more diagrams contributing to the three-point energy correlator of QCD jet that are not mediated by a virtual gluon.
Furthermore, for the diagrams that {\it are} mediated by a virtual gluon, the splittings of $g^*\to gg$ and $g^*\to q\bar{q}$ are not distinguishable if no flavor tagging criterion is imposed, and their contributions to the $\cos2\phi$ correlation have opposite signs to each other~\cite{Chen:2020adz, Hara:1988uj}.
As a result, the $\langle \xi_j \rangle$ is rather small. The analytic formula in the collinear limit is given by Eq.~(3) of~\cite{Chen:2020adz}.
For an active fermion number $n_f = 5$, $\langle \xi_j \rangle$ is $-0.01$ for quark jets and $-0.006$ for gluon jets.

\begin{figure}[htbp]
\centering
\includegraphics[scale=0.21]{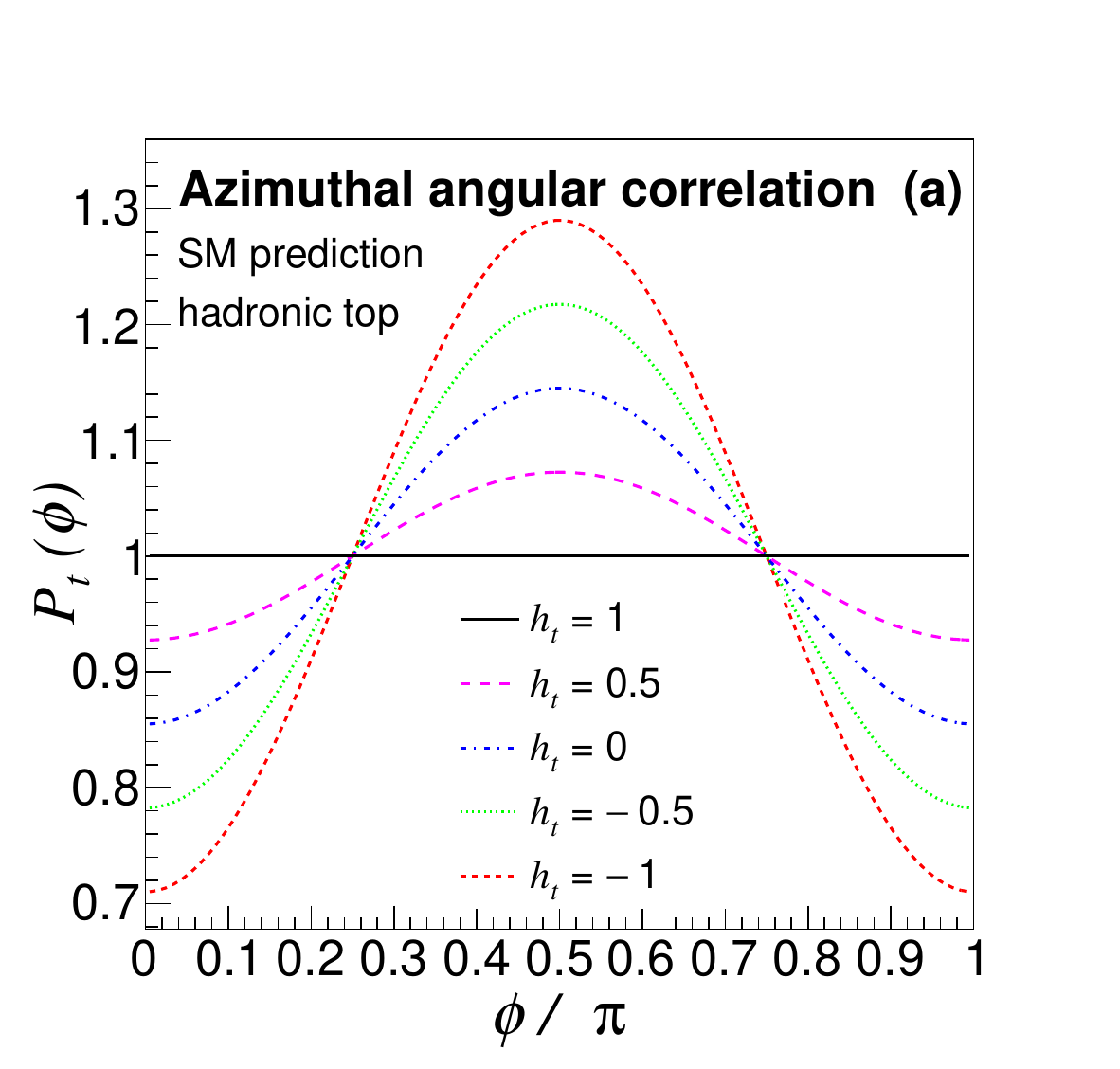}
\includegraphics[scale=0.21]{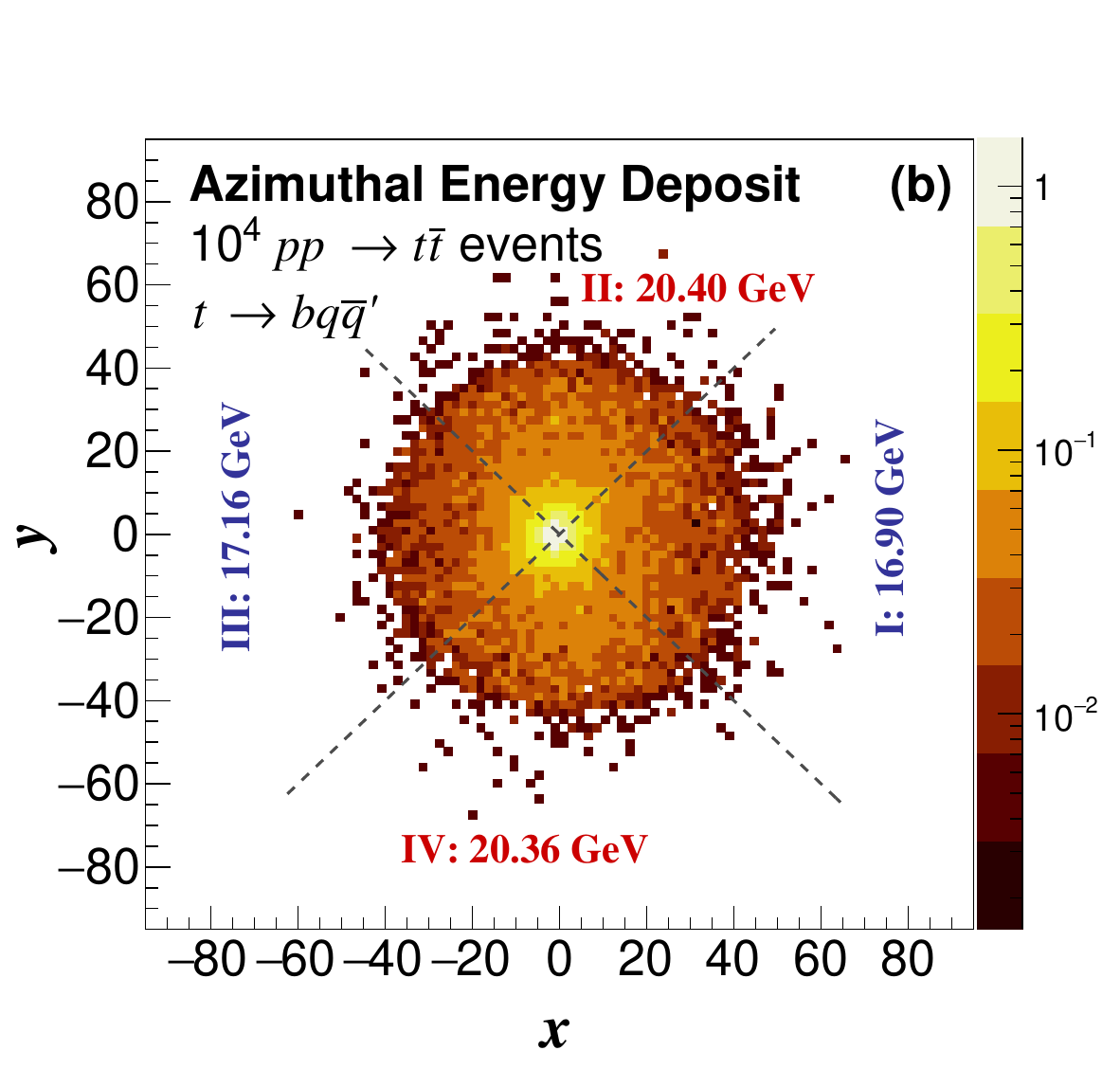}
\caption{(a) Azimuthal angular correlation in the decay of boosted top quark for different values of top longitudinal polarization $h_t$. (b) The transverse momentum distribution of $W$ decay products in the azimuthal plane of $W$ frame, viewed from the $z$ direction in Fig.~\ref{fig:frame}. The accumulated transverse momentum, averaged over $10^4$ events, has been indicated in each quadrant.}
\label{fig:phi distribution}
\end{figure}

\emph{Phenomenological implication.}---Here, we discuss a few applications of the proposed azimuthal correlation for exploring some aspects of top quark phenomenology at the LHC.

On the one hand, the $h_t$ dependence of $\langle \xi' \rangle$ in Eq.~\eqref{eq:phi distribution} enables the measurement of the longitudinal polarization of the top quark, without the need to reconstruct the top rest frame.
For example, in Fig.~\ref{fig:phi distribution}(a), we show the azimuthal correlation for a few different values of $h_t$ in the boosted limit.
The top quark polarization can give clues about its production mechanism, which is useful in testing the SM and searching for NP.
For example, in the QCD production of $t\bar{t}$ pairs, the inclusive top (anti)quark  should be unpolarized because QCD preserves parity symmetry, while in the single top production through electroweak interaction, {\it i.e.}, the $s$- or $t$-channel single top and $Wt$ productions, the top quark should be predominantly left-handed
because the charged current interaction is purely left-handed in the SM.
In various NP models, top quarks can be produced with various degrees of polarization~\cite{Berger:2011hn}.
Hence, the measured value of $\langle \xi' \rangle$ can help discriminate NP models. Below, we show how to construct such an experimental observable in hadronically decayed tops.

Even though we only performed a leading order calculation in the analysis, the $\cos2\phi$ correlation arises from the $W$ boson polarization, which is robust against perturbative QCD correction~\cite{Do:2002ky} and parton showering. In reality, we need to take the latter into account by defining an infrared (IR) safe observable. Note that the energies of $W$ decay products are not correlated with the azimuthal angle $\phi$, and therefore Eq.~\eqref{eq:phi distribution} can directly translate into energy distribution in the transverse plane of the $W$ frame,
\begin{align}
\frac{\dd E}{\dd \phi} = 
\frac{E_{\rm tot}}{2\pi} \left( 1 + \langle \xi' \rangle \cos2\phi \right) ,
\quad
\phi \in [0, 2\pi),
\label{eq:energy phi distribution}
\end{align}
where $E$ can also be taken as the transverse momentum in the $W$ frame, which is equally IR safe, and we have extended $\phi$ to $[0, 2\pi)$.

The $\cos2\phi$ distribution leads to an asymmetry of azimuthal energy deposition between the regions with $\cos2\phi > 0$ and $\cos2\phi < 0$, which divides the transverse plane into four quadrants, as shown by the two dashed diagonal lines in Fig.~\ref{fig:phi distribution}(b). This consideration motivates the following method to extract the coefficient $\langle \xi' \rangle$ that is suitable in experimental analysis:
\begin{enumerate}
\item[(1)] construct the top jet and its four-momentum $p_t^{\mu}$;
\item[(2)] use jet substructure technique with $b$ tagging to reconstruct the $b$ subjet with its four-momentum $p_b^{\mu}$;
\item[(3)] determine the $W$'s four-momentum $p_W^{\mu} = p_t^{\mu} - p_b^{\mu}$;
\item[(4)] construct the $W$ frame coordinate system ($x$-$y$-$z$) as in Fig.~\ref{fig:frame}, {\it i.e.}, $z$ along $\bm{p}_W$ and $y$ along $\bm{p}_b \times \bm{p}_W$; and
\item[(5)] remove the particles in the $b$ subjet and determine the energy distribution of the rest of top quark jet in the transverse plane ($x$-$y$).
\end{enumerate}
This method does not require identifying the quarks or subjets from $W$ decay. As a demonstration, in Fig.~\ref{fig:phi distribution}(b) we show the transverse energy deposit distributed in the azimuthal plane of $W$ frame, which is the average of $10^4$ hadronically decayed top quarks with $p_T \in (500, 600)~{\rm GeV}$ from the $t\bar{t}$ pair production in proton-proton collision at $\sqrt{s} = 13~{\rm TeV}$. The decayed events are generated with {\tt MG5\_aMC@NLO 2.6.7}~\cite{Alwall:2014hca} at leading order and passed to {\tt Pythia 8.307}~\cite{Bierlich:2022pfr} for parton showering, 
with full initial and final state radiations. Since hadronization is not correlated with the azimuthal distribution, it will not change the IR-safely defined azimuthal asymmetry. A similar argument also holds for the effect of underlying events that cancel in the asymmetry observable.
The anti-$k_T$ algorithm~\cite{Cacciari:2008gp} implemented in {\tt FastJet 3.4.0}~\cite{Cacciari:2011ma, Cacciari:2005hq} is used for the jet analysis, with a radius parameter $R = 1.0$ for finding the top jets and $R= 0.2$ for reclustering the top jets and identifying the $b$-tagged subjets. The energy deposits in the four quadrants are denoted as $E_1, \cdots, E_4$, sequentially, which have been indicated in Fig.~\ref{fig:phi distribution}(b). Evidently, there are more energy deposits in the $y$ direction, perpendicular to the $tbW$ plane, than the $x$ direction, which is parallel to the $tbW$ plane. Then we have
\begin{align}
\langle \xi' \rangle 
= \frac{\pi}{2} \cdot \frac{(E_1 + E_3) - (E_2 + E_4) }{(E_1 + E_3) + (E_2 + E_4) }.
\label{eq:exp-def}
\end{align}
This gives $\langle \xi' \rangle = -0.141 \pm 0.016$ in the simulated $t\bar{t}$ events, which agrees well with analytic calculation in Eq.~\eqref{eq:xi analytic} for top helicity $h_t = 0$. 
The quoted uncertainty is only of statistical origin, which is the dominant uncertainty in asymmetry observables~\cite{ATLAS:2019fgb, CMS-PAS-SMP-21-002}.
When using the same event selection criteria as in Ref.~\cite{arXiv:2205.02817}, which yields $17\, 261$ boosted $t\bar{t}$ events at the LHC Run-2 with $139~\ivfb$ integrated luminosity, we obtain an uncertainty $\delta{\langle \xi' \rangle} = 0.012$. Hence,  the azimuthal correlation can already be observed with the Run-2 data. Since $\delta{\langle \xi' \rangle} \propto 1/\sqrt{N_{\rm events}}$, we can project an uncertainty of 0.008 for $300~\ivfb$ at the LHC Run-3 and 0.0026 for $3000~\ivfb$ at the High-Luminosity LHC~\cite{Apollinari:2120673}. It is evident that the LHC data allow the precision measurement of such azimuthal correlation.

On the other hand, the hadronically decayed boosted top quark may well be clustered into a single jet by some jet algorithm, which may be contaminated by some QCD jet background events. To have a precision measurement of the top event rate, it is necessary to distinguish top jets from QCD jets.
Here, instead of constructing an event-by-event top tagger against QCD jets, we propose a simpler ``tagger" that acts on the whole ensemble of boosted top candidates to determine the fraction of top quark events.
In this ensemble, one can 
first measure the azimuthal asymmetry coefficient $\xi_0$ following the same strategy discussed above. This $\xi_0$ is not the same as the one for pure top quark events, as given in Eq.~\eqref{eq:xi analytic}, but is for a mixture of top and QCD jet events.
Then, if the top quark events account for a fraction $\delta_t$ of the whole ensemble, we should have
$\xi_0 =  \delta_t \, \langle \xi' \rangle + (1 - \delta_t) \, \langle \xi_j \rangle$,
from which we can determine $\delta_t = \left( \xi_0 - \langle \xi_j \rangle \right) / \left(\langle \xi' \rangle - \langle \xi_j \rangle \right)$,
where $\langle \xi_j \rangle$ is obtained by averaging over the light quark and gluon jet contributions and only depends on their relative fraction 
in the boosted QCD jet events. As an example, for single top quarks produced via $s$-channel SM-like heavy resonance $W^\prime$ with a mass $> 1~{\rm TeV}$, $\langle \xi' \rangle \sim -0.29$,
while the magnitude of $\langle \xi_j \rangle \lesssim 0.01$. As long as the top quark yield is not more than an order of magnitude smaller than the QCD jet background rate, $\delta_t$ can be precisely determined from the measurement of $\xi_0$ to constrain the parameter space of this NP model, such as the $W^\prime$-$t$-$b$ coupling strength. We leave a more detailed phenomenological study for future publication.

\emph{Conclusion.}---In this Letter, we proposed a novel substructure observable in the boosted top quark jet based on the azimuthal correlation between the $t\to bW$ and $W\to f\bar{f}'$ decay planes. The boosted top quark decays into a $W$ boson with a linear polarization, which results in a $\cos2\phi$ azimuthal correlation and translates into an energy deposition asymmetry in the azimuthal plane. Such linear polarization is not present in the top rest frame but only emerges under the boost as a result of mixing with other polarization parameters.
We have also demonstrated that such correlation can be used to either measure the longitudinal polarization of a boosted top quark for testing the SM and probing NP or distinguish a boosted top quark from the QCD jet background. 

\vspace{3mm}
\emph{Acknowledgments.}---This work is in part supported by the U.S.~National Science Foundation
under Grant No.~PHY-2013791. C.-P.~Y. is also grateful for the support from the Wu-Ki Tung endowed chair in particle physics.

\bibliographystyle{apsrev}
\bibliography{reference}

\end{document}